\documentclass[12pt,preprint]{aastex}



\def\deg{\hbox{deg}}
\def\Mpc{\hbox{Mpc}}

\lefthead{Huang et al.~}

\begin{document}
\title{The Hawaii-AAO $K$-band Galaxy Redshift Survey --- Paper I:
The Local $K$-band Luminosity Function
}
\author{J.-S.\ Huang,$\!$\altaffilmark{1,}~$\!$\altaffilmark{3,}~$\!$\altaffilmark{5}
K. Glazebrook,$\!$\altaffilmark{2,}~$\!$\altaffilmark{4}
L.\,L.\ Cowie,$\!$\altaffilmark{3}
C.\ Tinney$\!$\altaffilmark{4}
}

\altaffiltext{1}{Harvard-Smithsonian Center for Astrophysics, 60 Garden Street,
Cambridge, MA02138}
\altaffiltext{2}{Department of Physics \& Astronomy,
The John Hopkins University, 3400 N. Charles Street, Baltimore, MD21218}
\altaffiltext{3}{Institute for Astronomy, University of Hawaii, 2680 Woodlawn,
Honolulu, HI96822}
\altaffiltext{4}{Anglo-Australian Observatory, P.O. Box 296, Epping NSW 1710,
Australia}
\altaffiltext{5}{Max-Planck-Institut f\"{u}r Astronomie, K\"{o}nigstuhl 17,
D-69117 Heidelberg, Germany}


\begin{abstract}

  We present the $K$-band local luminosity function derived froma sample of
1056 bright ($K<15$) $K$-selected galaxies
from the Hawaii-AAO $K$-band
redshift survey.
The Hawaii-AAO $K$-band redshift survey covers
4 equatorial fields with a total area of 8.22 $\deg^2$. We derive both
the non-parametric and Schechter luminosity function from our data,
and determine $M^*(K)=-23.70\pm0.08+5\log_{10}(h)$, $\alpha=-1.37\pm0.10$ and
$\phi^*=0.013\pm0.003~h^3~\Mpc^{-3}$ for a universe with
$\Omega_m=0.3$ and $\Omega_{\Lambda}=0.7$.
We also measure the $K$-band luminosity function for the early- and later-type
galaxies from our morphologically classified subsample.
It appears that later-type galaxies have a fainter $M^*$ and a steep
slope, while early-type galaxies have a much brighter $M^*$ and a
quite flat slope in their $K$-band luminosity functions. This is
consistent with what have been found in optical type dependent luminosity
function.
The $K$-band luminosity density derived using our luminosity function is now
measured at a similar redshift depth to optical luminosity densities in the
SDSS redshift survey. It is 2 times higher than the previous measurement from
the shallower 2MASS sample and resolves the previously reported discrepancies
between optical and near-IR luminosity densities.

\end{abstract}


\section{Introduction}

   The galaxy luminosity function is an important quantity in the study of
galaxy evolution and formation. Traditionally,
galaxy luminosity functions have been derived in optical bands.
It is now clear
that extragalactic studies in optical bands suffer
several systematic uncertainties and complexities
compared to those using near infrared bands. In particular,
dust extinction has little effect on $K$-band
magnitudes; the K-correction in the $K$-band is a much smaller and better
understood  quantity than in the optical bands, and it is independent of galaxy
spectral types for $z<1$. Because of this it is easier to detect a
high redshift elliptical galaxy in an infrared band than in optical bands.
A galaxy's near infrared luminosity is also a good tracer of its
stellar mass independent of spectral type \citep{col01,gla02}.
Recent theoretical studies
show that the $K$-band galaxy luminosity function is
a powerful constraint on galaxy formation theory \citep{bau98, kau98}.

   Most near infrared surveys, however, have been modest in size
due to the small size of available infrared detectors.
The advent of large format infrared array detectors has made
possible a variety of wide-field near-infrared surveys,
ranging from several at the 10 deg$^2$-level \citep{gar96, hua97}
up to the largest of them all, the 2 Micron
All Sky Survey (2MASS) \citep{skr97}. Obtaining optical
redshifts for a $K$-selected sample, however, is difficult because
the wide range in optical-infrared colours results in the requirement
for a wide range of exposure times to acquire optical spectroscopy.
In particular, very long exposures are required to secure redshifts
for the reddest objects. Because of this there are only a
few $K$-band luminosity functions available. The early
$K$-band luminosity functions were derived from small size samples with number
of galaxies ranging from 100 to 500
\citep{mor93, gla95, gar97, szo98, lov00}.
After the second incremental release of 2MASS data, 
two teams \citep{col01,koc01}
used the overlap between the 2MASS Extended
Source Catalog (http://pegasus.phast.umass.edu) and 
two existing optical redshift databases, CfA2 \citep{del88} and 
2dFGRS (http://www.mso.anu.edu.au/2dFGRS),
to obtain very large ($>4000$), if
very shallow ($K<13$), $K$-selected redshift samples to derive the local
$K$-band luminosity function. However, the mean redshift for these
samples is shallow, 0.025 for the CfA sample\citep{koc01} and 0.05
for the 2df sample\citep{col01}.

   The observational goal of the Hawaii-AAO $K$-band redshift survey was
to obtain a large medium-deep $K$-selected galaxy sample with
redshifts and optical-infrared colors.
Both optical ($B$ and $I$) and near-infrared ($K$)
images were taken at Mauna Kea Observatory with total infrared coverage of
totaling 8.22 $\deg^2$ \citep{hua97}.
The spectroscopic observations were carried out on the Anglo-Australian
Telescope (AAT) with the Two Degree Field facility (2dF).
In this paper we report $K$-band luminosity functions derived from
a sub-sample of 1056 bright ($K<15$) galaxies.
The median redshift of this sample is 0.138 with the redshift distribution tail extending to z=0.5.
In this paper, we adopt the
$\Omega_m=0.3$ and $\Omega_{\Lambda}=0.7$ and $H=100~\rm km~s^{-1}~Mpc^{-3}$ cosmology model.
A comparison of the $K$-band luminosity functions derived
using different cosmology
models is also presented. We briefly summarize the observation
and data reduction in $\S$2; in $\S$3, we present the $K$-band luminosity
functions; in $\S$4, we compare our luminosity function with other's and
discuss the difference; we summarize our results in $\S$5.

\section{Observations}

   \citet{hua97} have described in detail the acquisition and processing of
our $B$-,$I$- and $K$-band imaging.
In brief, the imaging survey was carried out at
Mauna Kea Observatory and represents over 70 nights of time on the University
of Hawaii 88-inch and 24-inch telescopes. The survey consists of
four equatorial fields covering a total area of 8.22deg$^2$.
The optical and near-infrared
limiting magnitudes are $B_{limit}=22$, $I_{limit}=20.5$,
and $K_{limit}=16.0$.
The spectroscopic observations for the $K$-selected sample were carried out
at the AAT with 2dF, a multi-fiber spectrograph
that can observe 400 targets within a 2 degree field of view. These observations
took approximately 10 clear nights spread out over the period 1997--1999.
The exposure time for a spectroscopic observation was determined by
the $I$ magnitude of each object, for objects with $I<18.5$ ($I-K<2.5$ the
$K=16$ limit) 1 hour exposure times sufficed, for the redder objects
($18.5<I<21$) exposure times of up to 4 hours were used.
Weather variations meant that some 2dF observations were
not as deep as others. When a redshift was not secured at a first
attempt the object was flagged for re-observation in a later
2dF configuration.
Multiple 2dF configurations were observed on
each field to ensure maximal completeness.

Our $K<15$ sample is highly complete in redshift.
 From a total of 1201 objects imaged with $K<15$, redshifts were
obtained for 1056 galaxies. A further 59 objects turned out to be
stars, leaving only 86 objects without identification. These objects
are assumed to be galaxies for the purposes of incompleteness correction.
Fig.1 shows the redshift distribution
of the sample. The median redshift for
the $K\le15$ sample is 0.136.
We calculate absolute magnitudes with a
$\Omega_m=0.3$ and $\Omega_{\Lambda}=0.7$ cosmology and use recently published
K-correction due to \citet{man01}. K-corrections have also
be tested using spectral evolution models which best fit the observed
colours \citep{PEGASE}. No significant difference was
found in either the K-corrections or the resulting luminosity functions.

\section{The Luminosity Function}

   We use the Stepwise Maximum Likelihood (SWML) \citep{efs88} to
derive the non-parametric luminosity function, and use the STY method
\citep{san79} to fit the Schechter function to the data.
After obtaining $M^*$ and $\alpha$ by fitting Schechter function to our
 data,
we use the minimum variance estimator \citep{dav82} to determine the
$\phi_*$.
Both STY and SWML methods are based on the maximum likelihood
principle, and therefore only valid
for a complete galaxy redshift sample. Correction for our sample's
(small) incompleteness is therefore needed.  \citet{zuc94, lin96, lin99}
have shown that a {\em weighted STY} method can
be used to derive a luminosity function from an incomplete sample.
In the weighted STY method, if a galaxy in the sample
has no redshift,
similar galaxies with redshifts (ie. similar magnitude, or similar colour,
or both) receive increased weight in calculating the likelihood
to represent the galaxy without a redshift,

\begin{center}
\[
P_i=\left(\frac{\psi(L_i)}{\int_{L_{min}(z_i)}^{\infty}\psi(L) dL}\right)^{w_i}
\]
\end{center}

Here the $P_i$ is the probability for galaxy i at redshift $z_i$,
$\psi(L)$ is the luminosity function, and $w_i$ is the weight for
galaxy i. $w_i=1$ for a complete sample, otherwise $w_i\ge1$.

   In our case only 7\% of targets do not have redshifts, so such
a correction does not make a large difference. We fit Schechter
functions to the sample both with, and without, correction.
We follow \citet{lin96, lin99} in estimating the weight
function for each galaxy with redshift in the sample. We divide the
sample into apparent magnitude and color $I-K$ bins, and the weight $w_i$
assigned to galaxy i with redshift $z_i$ is the ratio of total number of
galaxies over the total number galaxies with redshifts in the bin where
the galaxies i is. Since the incompleteness of the 4 fields varies, we have
to calculate $w_i$ for galaxies in each field separately.
Table~1 summarizes the result of the STY fitting with 3 different cosmological
models, and Fig.~2 shows the non-parametric K-band luminosity function and 
the Schechter function in the
$\Omega_m=0.3$ and $\Omega_{\Lambda}=0.7$ universe.

  As shown in Table~1, the parameters of the Schechter function
depend on the adopted cosmological models. This is because the median
redshift is 0.138, where the effect of the geometry for
the adopted cosmology model appears in calculating distance,
and the change of $\alpha$ is due to the correlation between $M^*$ and
$\alpha$.
There is a slight difference between the luminosity functions derived
with correction and without correction. The $M^*$
calculated with correction is slightly brighter
than that calculated without correction. This is because
most of the galaxies without redshifts are near the $K=15$
limit and are either bright galaxies at relatively high
redshifts comparing with rest of the sample, or local lower luminosity
galaxies. Similar galaxies with redshifts in the sample are assigned more weight.
The consequence is to cause $M^*$ to become somewhat  brighter and the faint end stope
slightly steeper after correction.

   Using the higher spatial resolution $I$-band images,
\citet{hua98} was able to make a
morphological classification for the very bright galaxies ($K \le 14$).
Among 225 galaxies with $K \le 14$ in the sample,
\citet{hua98} classified 111 galaxies
as E/S0. Most of the remaining 114 galaxies are spirals, and only a few
with peculiar morphologies are possible mergers or irregular galaxies.
Therefore, we are able to derive the luminosity function of early type
(E/S0) and later type (Spiral and Irregular galaxies) for
the $K \le 14$ sample. Fig~3 shows the the $K$-band luminosity function
for both types with the best fitting Schechter functions.
The $K$-band luminosity function for the early type galaxies
has a bright $M^*$ of $M^*=-23.56\pm0.26$ and a flat slope
$\alpha=-1.04\pm0.31$, while that for the later type galaxies has
a much fainter $M^*$ of $M^*=-23.28\pm0.28$ and a steeper slope
$\alpha=-1.42\pm0.31$.

\section{Discussion}

   Recently, several $K$-band luminosity functions were derived either directly from
$K$-band surveys \citep{gla95, gar97, szo98}, or
from optical samples with $K$-band images \citep{mor93, lov00}.
These luminosity functions are roughly in agreement
with each other, given that most samples are small in size and
have large statistical uncertainty.
To compare the $K$-band
luminosity functions in different cosmology
models, \citet{col01} convert all existing $K$-band luminosity functions
to those in the $\Omega_m = 0.3$ and $\Omega_{\Lambda}=0.7$ cosmology model.
Table 2 shows the parameters of Schechter function for all $K$-band
luminosity functions, including those we adopted from \citet{col01} and ours.
We also list sample size and limiting magnitude for each survey.

   By comparing the Schechter functions, our luminosity function has
a brighter $M^*$ and a steeper slope, only those of
\citet{szo98} are close to ours. The rest of the luminosity functions
have a $M^*$ at least 0.3 magnitude fainter than ours and a flat slope
($\alpha\sim -1$), including two 2MASS luminosity functions
\citep{col01,koc01} and that
of \citet{gar97} who had a similar sample to ours.
For more accurate comparison, we plot the luminosity
functions against each other in Fig.~4.
Fig.~4 shows that our luminosity function is higher and steeper
than the 2MASS luminosity functions and that of \citep{gar97}. This
is in agreement with the comparison in Schechter functions that
the 2MASS luminosity functions have fainter $M^*$, lower $\phi^*$,
and flat $\alpha$.

  There are two different important aspects between our sample and the 2MASS
sample\citep{col01,koc01}: different ways of measuring total
magnitudes for galaxies and different redshift ranges for both sample.
Each of them could cause the difference in deriving luminosity function.
If we assume that both photometric methods measure true total magnitudes for
$K$-selected galaxies, the difference between both $K$-band luminosity functions
implies a change of luminosity function at different redshift ranges.
Since our sample covers a wider redshift range, we are able to derive
the $K$-band luminosity function in the lower redshift bin.
This will allow us to test if we are able to reproduce the results from
the 2MASS sample using our lower redshift bin sample. The limiting
magnitude in both \citet{col01,koc01} samples are around $K\sim13$,
which is too bright for our sample to be statistically significant.
However we can adopt a redshift limited subsample which approximates
their redshift range, it will simply extend $\sim$ 2 magnitudes 
further down the
luminosity function. Thus we set the limiting
redshift $z<0.1$, to approximate the redshift range to those of the 
2MASS sample
while allowing us to have enough galaxies to derive the luminosity function.
Fig.~4 shows that our luminosity function for the subsample with $z<0.1$
matchs those luminosity functions of \citep{col01,koc01} very well.
We also fit the Schechter function to our subsample with $z<0.1$, and
obtain that $M^*=-23.10\pm0.15$,
$\alpha=-0.93\pm0.16$, and $\phi^*=0.012\pm0.004~Mpc^{-3}$ consistent
with those derived using the 2MASS samples.
This implies that the normlisation of the luminosity function 
is a function of redshift,
even at lower redshifts, and that this explains the apparent discrepancy 
in published
estimates without needing to invoke systematic errors in deriving
total magnitudes in both samples.

  The steep slope of our $K$-band luminosity function is actually consistent
with those of optical luminosity functions
obtained in several current wide-field
optical surveys. \citet{fol99} drived a
$B$-band luminosity function with $\alpha=-1.28\pm0.05$ using a large
redshift sample obtained in the 2df Galaxy Redshift Survey (2dFGRS).
\citet{bla01}, and the SDSS team obtained the optical luminosity
functions in $u^*$, $g^*$, $r^*$, $i^*$, and $z^*$ bands,
and the $\alpha$ for these luminosity functions in the
5 bands are $-1.38$, $-1.26$, $-1.20$, $-1.25$, and $-1.24$ respectively.
\citet{col01} used the average color to transfer the SDSS $z^*$-band
luminosity function to a $K$-band luminosity function with $M^*=-23.67$,
$\phi^*=0.0127$, and $\alpha=-1.24$, which are very close to the
parameters of our luminosity function.
We would like to point out that the peak of the redshift distribution
for the SDSS sample is about 0.1, much closer to ours than to those of
the 2MASS sample \citep{col01,koc01}. The implies that the normalization
of the luminosity function is a function of redshift, even at lower redshifts,
and that this explains the apparent discrepancy in published estimates
without needing to invoke systematic errors in deriving total magniutdes
in both sample.

   Before  the 2MASS sample was available,
there were no K-selected samples with morphological classification, hence no
type-dependent K-band luminosity functions. \citet{lov00} derived
K-band luminosity function for the emission line galaxies (ELG) and also for
the galaxies without emission lines (Non-ELG). He found that the $M^*$
of the K-band Luminosity function for ELG is one magnitude fainter 
than that for non-ELG. 
For the first time, \citet{koc01} was able to derive
the $K$-band luminosity function for both early and later type galaxies.
We can compare ours with those of \citet{koc01}. Since
both samples have very bright limiting magnitudes,
$K<11.2$ for \citet{koc01} and $K<14$ for our sample,
the effect of the geometry for the adopted
cosmology models cannot make
any significant difference in calculating absolute magnitudes
at such a low redshift.
We can compare the two results directly.
In Table~3 we list the $M^*$ and $\alpha$ derived
from both our $K<14$ sample and the 2MASS sample
\citep{koc01}. Our $K$-band luminosity function for early type
galaxies has a similar $M^*$ and $\alpha$ to those of \citet{koc01},
but the luminosity functions for later
type galaxies are more different: our $M^*$ is 0.3 magnitude brighter and our slope is much
steeper ($\alpha= -1.42$) than theirs ($\alpha=-0.87$). 
Steep slopes ($\alpha<-1$)
for later type galaxy luminosity functions
are also seen in the optical bands \citep{mar94, bro98, mar98, fol99}.

  The spectrum of the universal luminosity density is another way of
testing consistency between optical and infrared luminosity functions
\citep{wri01}.
\citet{dwe98} suggested
that the spectrum of the universal luminosity density can be
fitted by an average spiral
galaxy SED \citep{sch97}.  \citet{wri01}
used the average spiral galaxy SED model to fit the luminosity densities
derived from the SDSS luminosity function, and find that the model
predicts the luminosity densities at $J$ and $K$ bands should be
2.3 times higher than the
values derived from the 2MASS luminosity functions. In Fig.~5, we reproduce the
luminosity density plots by adding our points.
Our $K$-band luminosity density is about 2 times higher
than those derived from the 2MASS samples\citep{col01,koc01}, much closer
but still lower than what the model predicts. The luminosity density of our
subsample with $z<0.1$ is consistent with those from the 2MASS sample.
This is not surprising since they are derived using similar luminosity
functions.
Our luminosity density for the $K<15$ sample
is measured at a mean redshift of $z\sim 0.1$, comparable to SDSS
and 2dFGRS {\em optical} samples. However the 2MASS-(SDSS,2dFGRS) paired
catalogs are much shallower, $z\sim 0.05$, thus it seems that the most
likely explanation is that 2MASS is sampling a local underdensity in the
galaxy distribution. However it is now clear that
at a redshift $\sim 0.15$ there
is no broad mismatch in optical and IR luminosity densities.
We also notice that the model,
the average spiral galaxy SED \citep{sch97}  plus
a constant tail ($f_{\nu}\propto\nu^{-2}$) at short wavelength
\citep{dwe98, wri01}, does not fit the luminosity densities well, in detail, either.

\section{CONCLUSIONS}

   We present a large and highly complete large $K$-selected redshift sample
down to $K\le 15$, and use it to derive the
$K$-band luminosity function using both non-parametric and STY
methods. By comparing with previous $K$-band luminosity functions, our $K$-band
luminosity function has a significantly brighter characteristic luminosity and steeper
slope. The slope is the same as in the optical determinations.
The $K$-band luminosity density measured from our sample
is 2 times higher than those measured from the 2MASS redshift sample,
the largest local $K$-selected redshift sample. We argue
that this deeper survey, of comparable depth to the optical and SDSS surveys,
is more strictly comparable in luminosity density and that we have in
fact resolved the discrepancy of the 2MASS survey. We are also able
to re-produce the the 2MASS $K$-band luminosity function using similar
sunsample with $z<0.1$.

   We also derive the $K$-band luminosity functions for both early and later type
galaxies.
The early type galaxies have a bright $M^*$ and a flat slop
$\alpha\sim -1$, while the later type galaxies have  a faint $M^*$ and
a steep slope $\alpha=-1.4$. A steep slope for later type galaxies is also
found in the current large optical redshift surveys.

\acknowledgements

   This research has been supported by University of Hawaii, AAO, Max-Planck
Institute for Astronomy, and the Smithsonian Astrophysical Observatory.
J.-S. Huang thanks both Max-Planck Institute for Astronomy and
AAO for their financial support during his visiting AAO.

\clearpage

\begin{figure}

\plotone{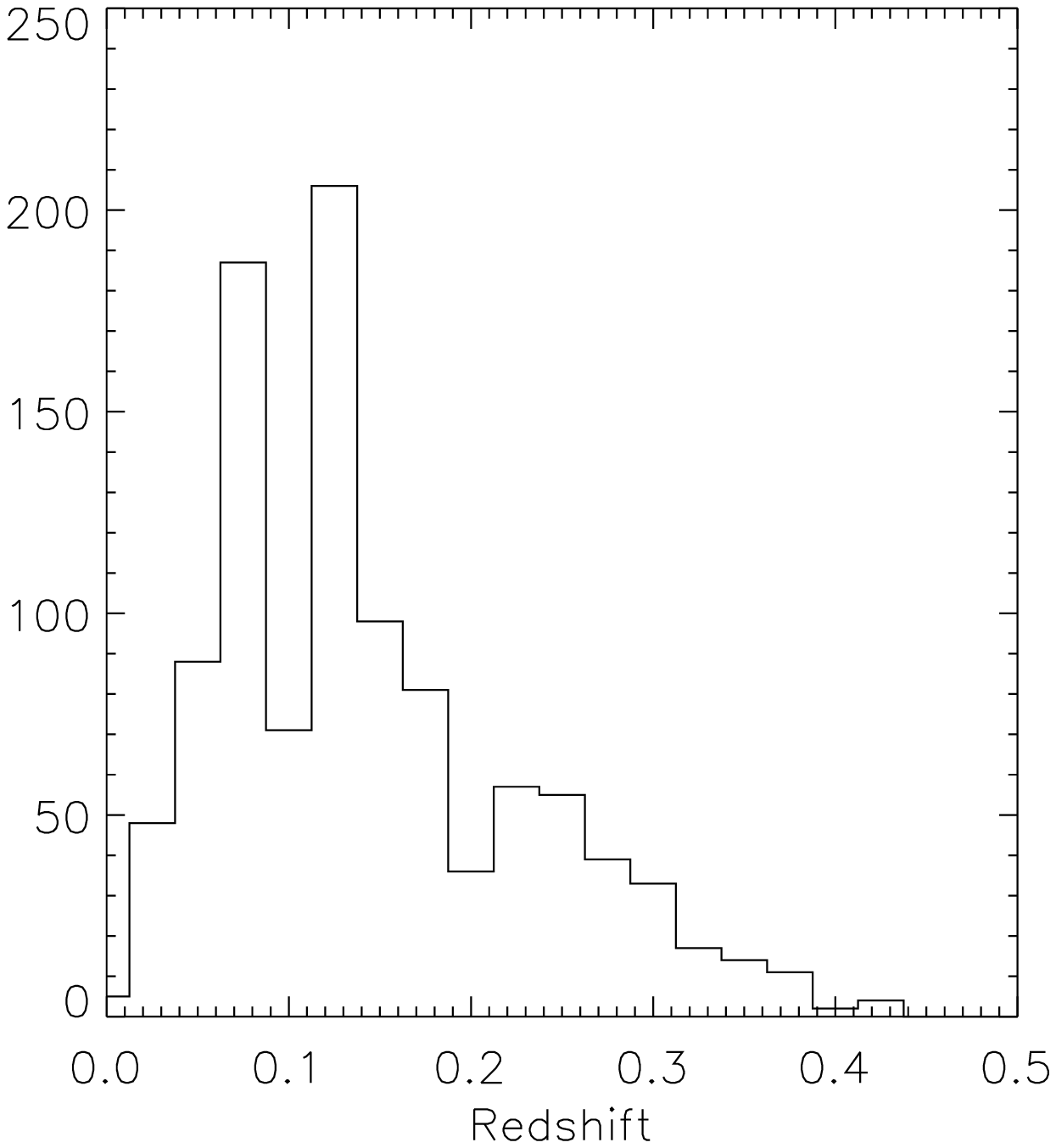}
\caption{ The redshift distribution for the $K \le 15$ sample of
totaling 1056 galaxies.
The median redshift is 0.138, and the maximum redshift is 0.57. \label{fig1}
}
\end{figure}

\clearpage
\begin{figure}
\plotone{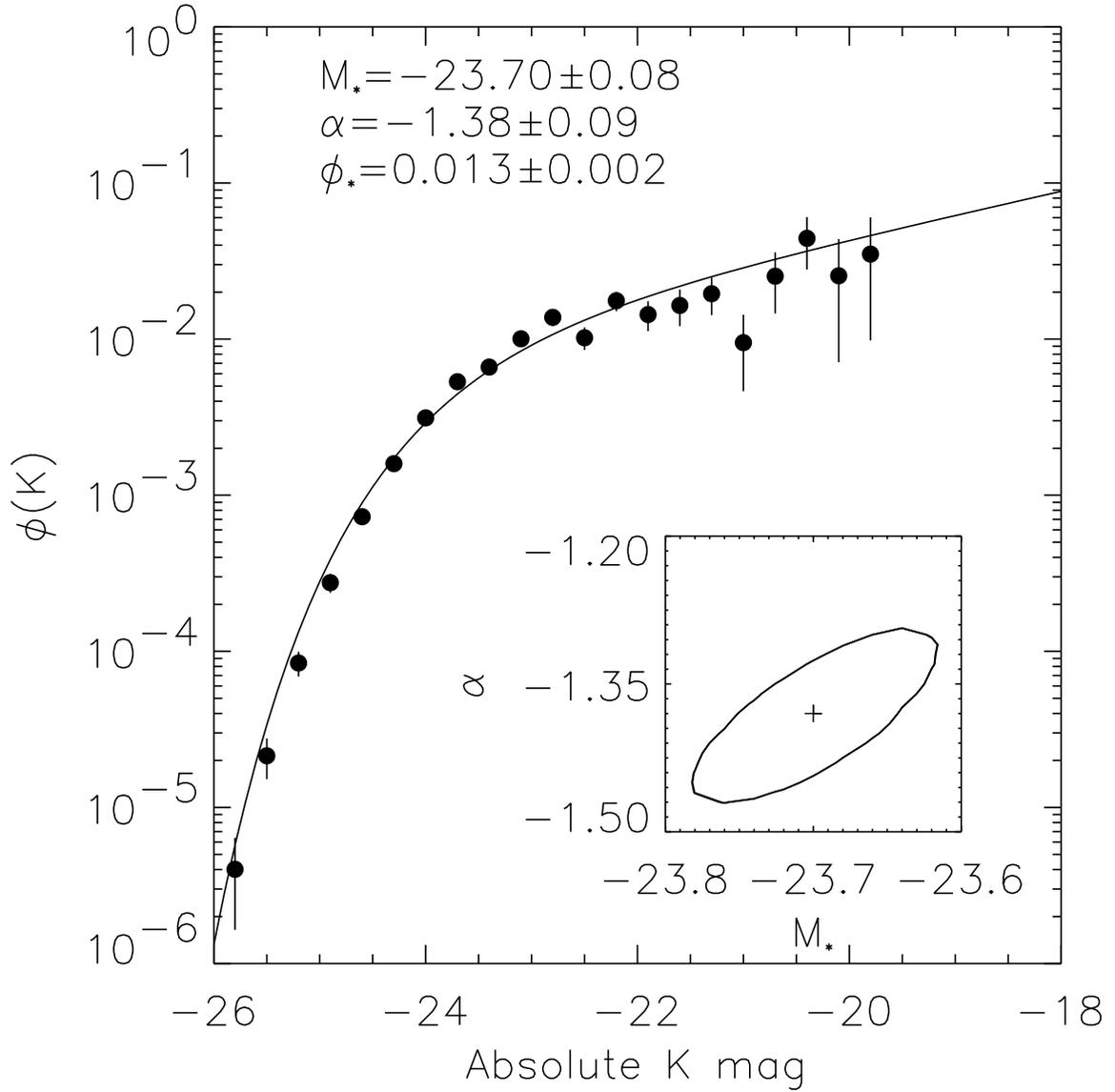}
\caption{The $K$-band luminosity function derived from our $K \le 15$ sample with
the best fit Schechter function. We also plot the 1$\sigma$ contour at
lower right corner. \label{fig2}
}
\end{figure}

\clearpage

\begin{figure}
\plotone{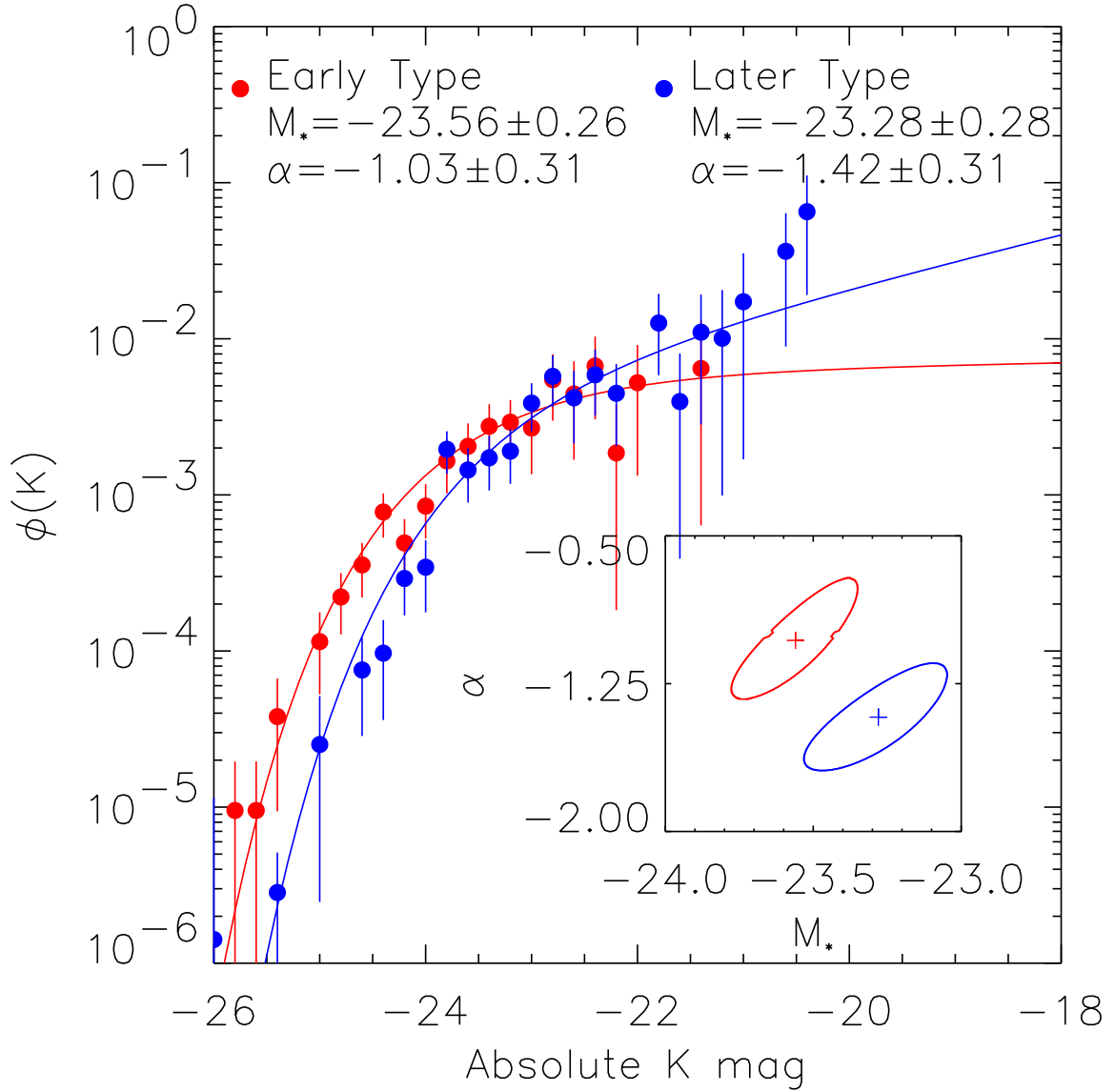}
\caption{
The $K$-band luminosity functions for the early and later type galaxies,
which are different from each other: the LF of the later
type galaxies has a much steeper slope end than that of the early type
galaxies; the LF for the early type galaxies has a bright $M_*$. \label{fig3}
}

\end{figure}

\clearpage

\begin{figure}
\plotone{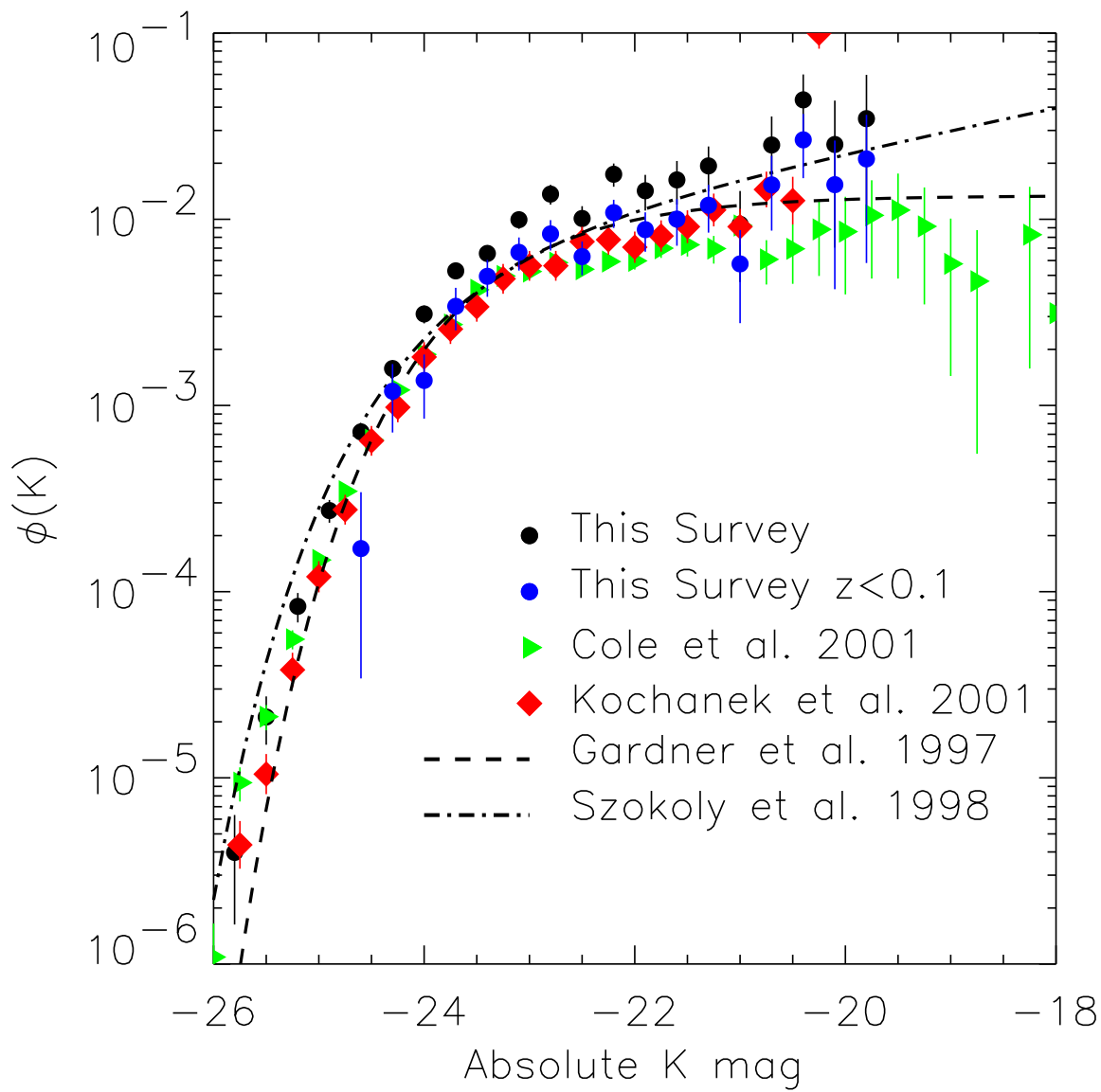}
\caption{
Our luminosity function is plotted against two 2MASS $K$-band luminosity
functions \citep{col01, koc01}, and those of \citet{gar97}
and \citet{szo98}. We also plot our luminosity function for $z<0.1$
which is consistent with those from the 2MASS sample.\label{fig4}
}
\end{figure}

\clearpage

\begin{figure}
\plotone{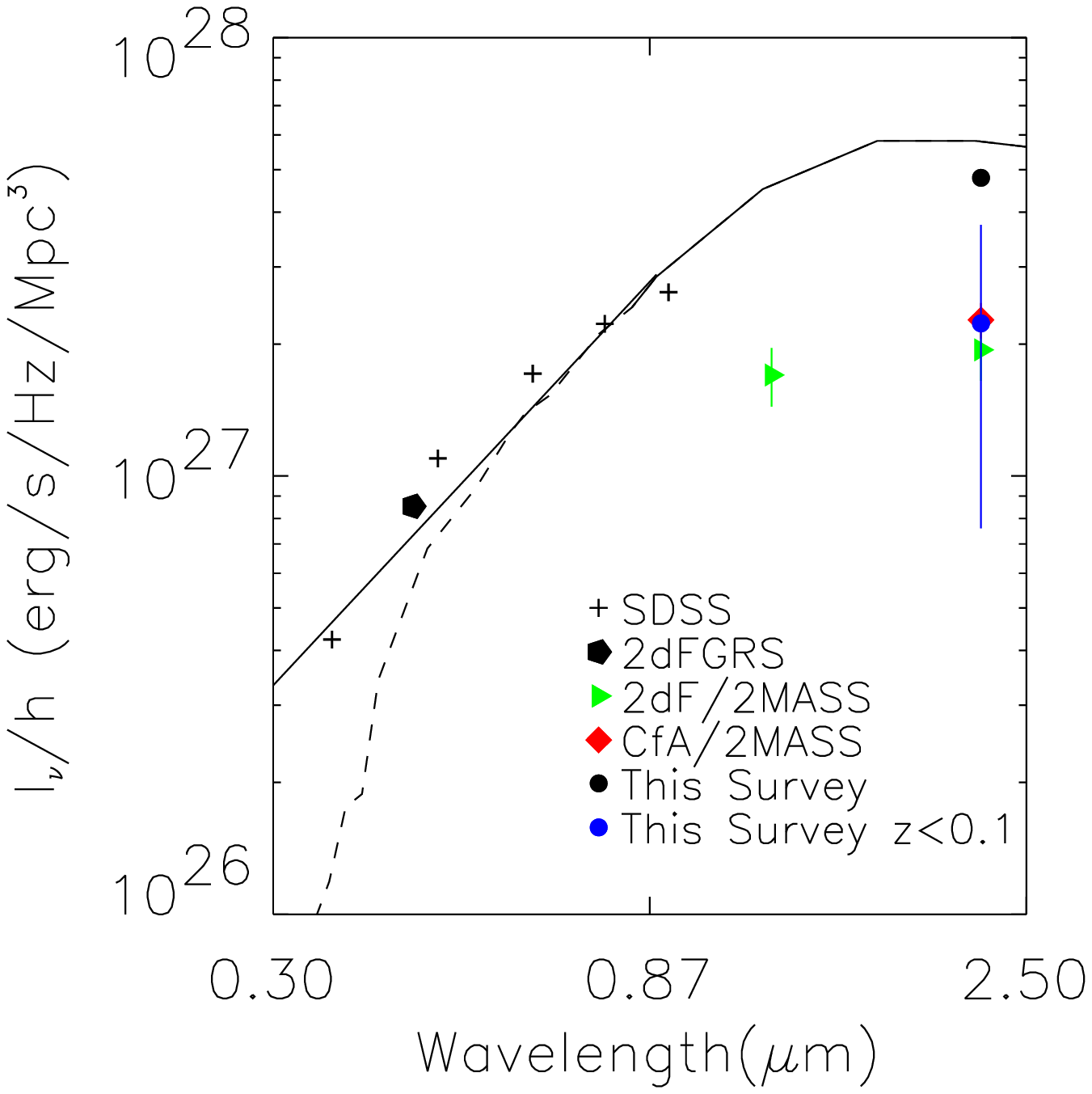}
\caption{
The optical and near-infrared luminosity densities derived with
$\Omega_{m}=0.3$ and $\Omega_{\Lambda}=0.7$ cosmology.
The optical luminosity
densities are measured in $u^*$,$g^*$, $r^*$, $i^*$, and $z^*$ by
the SDSS team\citep{bla01}, and in $B_J$ band from the 2dFGRS
\citep{fol99}. The J and $K$-band luminosity densities are derived
from the 2MASS luminosity functions \citep{col01, koc01},
and from our data. The luminosity density derived using our subsample at
the low redshift bin (blue filled dot) is consistent with the
2MASS luminosity density. An average spiral galaxy SED (solid line and
the one with an flat tail at short wavelength
($f_{\nu}\propto\nu^{-2}$, Dashed line) as a background model suggested by
\citet{dwe98} and \citet{wri01} are also plotted. \label{fig5}
}
\end{figure}

\clearpage

\begin{table*}[hbt]
{\scriptsize
\begin{center}
\centerline{\sc Table 1}
\vspace{0.1cm}
\centerline{\sc $K$-band Luminosity Function}
\vspace{0.3cm}
\begin{tabular}{lcccc}
\hline\hline
\noalign{\smallskip}
  & $M_*-5*\log(h)$ & $\alpha$ & $\phi_*$ & $l_{\nu}~^1$\cr
\hline
\noalign{\smallskip}
$\Omega_{m}=0.3~~\Omega_{\Lambda}=0.7 (uncorr^2)$ & -23.64$\pm$0.08 & -1.30$\pm$0.11 & 0.014$\pm$0.001& 4.05$\pm$0.13\cr
$\Omega_{m}=0.3~~\Omega_{\Lambda}=0.7 (corr^3)$ & -23.70$\pm$0.08 & -1.39$\pm$0.09 & 0.014$\pm$0.001& 4.79$\pm$0.16\cr
$\Omega_{m}=0.3~~\Omega_{\Lambda}=0.0 (uncorr)$ & -23.51$\pm$0.08 & -1.25$\pm$0.09 &0.017$\pm$0.002 & 4.14$\pm$0.44\cr
$\Omega_{m}=0.3~~\Omega_{\Lambda}=0.0 (corr)$ &-23.57$\pm$0.08 & -1.33$\pm$0.09 & 0.017$\pm$0.002 & 4.82$\pm$0.44 \cr
$\Omega_{m}=1.0~~\Omega_{\Lambda}=0.0 (uncorr)$ & -23.36$\pm$0.08 & -1.16$\pm$0.09 &0.020$pm$0.002 & 3.88$\pm$0.41\cr
$\Omega_{m}=1.0~~\Omega_{\Lambda}=0.0 (corr)$ &-23.41$\pm$0.08 & -1.25$\pm$0.09 & 0.020$\pm$0.002 & 4.44$\pm$0.40\cr
\noalign{\hrule}
\noalign{\smallskip}
\end{tabular}

\end{center}
\begin{flushleft}
$^1$~~Luminosity density in units of $10^{27} erg/s/Hz/Mpc^3$\\
$^2$~~LF derived without correction for incompleteness\\
$^3$~~LF derived with correction for incompleteness\\
\end{flushleft}
}
\label{tab1}
\end{table*}

\clearpage

\begin{table*}[hbt]
{\scriptsize
\begin{center}
\centerline{\sc Table 2}
\vspace{0.1cm}
\centerline{\sc Previous and Current $K$-band Luminosity Functions$^1$ }
\vspace{0.3cm}
\begin{tabular}{lcccccc}
\hline\hline
\noalign{\smallskip}
Sample & $M_{*}$ & $\alpha$ & $\phi_{*}~^2$& $l_{\nu}~^3$ & number & $m_{li
mit}$\cr
\hline
\noalign{\smallskip}
Mobasher et al.~1993 & -23.37$\pm$0.30 & $-1.00\pm0.3$ & 1.12$\pm$0.16 &
1.96$\pm$0.62 & 95&\cr
Glazebrook et al.~1995 & -23.14$\pm$0.23 & $-1.04\pm0.3$ & 2.22$\pm$0.53 &
3.19$\pm$1.07& 98 & $K \le 17.3$\cr
Gardner et al.~1997 & -23.30$\pm$0.17 & $-1.00\pm0.24$ & 1.44$\pm$0.20 &
2.36$\pm$0.48& 532& $K \le 15.0$\cr
Szokoly et al.~1998 & -23.80$\pm$0.30 & $-1.30\pm0.20$ & 0.86$\pm$0.29 &
2.90$\pm$1.27&110&$K \le 16.5$\cr
Loveday 2000 &  -23.58$\pm$0.42 & $-1.16\pm0.19$ & 1.20$\pm$0.08 &
2.86$\pm$0.94&345 & $b_j \le 17.15$\cr
Kochanek et al.~2001 & -23.43$\pm$0.05 & $-1.09\pm0.06$ & 1.16$\pm$0.10&
2.27$\pm$0.21& 3878& $K \le 11.25 $\cr
Cole et al.~2001 & -23.36$\pm$0.02 & $-0.93\pm0.04$ & 1.16$\pm$0.17&
1.94$\pm$0.29& 5683& $K \le 13.2$\cr
This paper & -23.70$\pm$0.08 & $-1.39\pm0.09$ & 1.30$\pm$0.20 &
4.79$\pm$0.16&1056 & $K \le 15.0$\cr
\noalign{\hrule}
\noalign{\smallskip}
\end{tabular}
\par

\end{center}
\begin{flushleft}
$^1$~~LF derived with $\Omega_{m}=0.3$ and $\Omega_{\Lambda}=0.7$.\\
$^2$~~$\phi_{*}$ in units of $10^{-2} h^3 Mpc^{-3}$.\\
$^3$~~Luminosity density in units of $10^{27} erg/s/Hz/Mpc^3$\\
\end{flushleft}
}
\label{tab2}
\end{table*}

\clearpage

\begin{table*}[hbt]
{\scriptsize
\begin{center}
\centerline{\sc Table 3}
\vspace{0.1cm}
\centerline{\sc $K$-band Luminosity Functions for early and later type galaxies}
\vspace{0.3cm}
\begin{tabular}{lcccc}
\hline\hline
\noalign{\smallskip}
{} &  \multicolumn{2}{c}{Early Type} & \multicolumn{2}{c}{Later Type} \cr
Sample & $M_{*}$ & $\alpha$  & $M_{*}$ & $\alpha$\cr
\hline
\noalign{\smallskip}
Kochanek et al.~2001 & -23.53$\pm$0.06 & $-0.92\pm0.10$ & -22.98$\pm$0.06&
$-0.87\pm0.09$\cr
This paper & -23.56$\pm$0.26 & $-1.03\pm0.31$ & -23.28$\pm$0.28 &$-1.42\pm0.31$\cr
\noalign{\hrule}
\noalign{\smallskip}
\end{tabular}

\end{center}
}
\label{tab3}
\end{table*}

\end{document}